\documentclass[12pt]{article}

\usepackage{natbib}
\usepackage{amssymb}
\usepackage{amsmath}
\usepackage{multirow}
\usepackage{latexsym}
\usepackage{epsfig}
\usepackage{graphicx}
\usepackage{epstopdf}
\usepackage[usenames]{color}
\usepackage{multicol}

\newcommand{\bfa}{\mbox{\boldmath $a$}}
\newcommand{\bfb}{\mbox{\boldmath $b$}}

\newcommand{\bfm}{\mbox{\boldmath $m$}}

\newcommand{\bfy}{\mbox{\boldmath $y$}}

\newcommand{\bfB}{\mbox{\boldmath $B$}}
\newcommand{\bfC}{\mbox{\boldmath $C$}}
\newcommand{\bfD}{\mbox{\boldmath $D$}}

\newcommand{\bfF}{\mbox{\boldmath $F$}}
\newcommand{\bfG}{\mbox{\boldmath $G$}}
\newcommand{\bfH}{\mbox{\boldmath $H$}}
\newcommand{\bfI}{\mbox{\boldmath $I$}}

\newcommand{\bfM}{\mbox{\boldmath $M$}}

\newcommand{\bfR}{\mbox{\boldmath $R$}}

\newcommand{\bfV}{\mbox{\boldmath $V$}}
\newcommand{\bfW}{\mbox{\boldmath $W$}}

\newcommand{\bfbeta}{\mbox{\boldmath $\beta$}}

\newcommand{\bfeta}{\mbox{\boldmath $\eta$}}

\newcommand{\bftheta}{\mbox{\boldmath $\theta$}}
\newcommand{\bfomega}{\mbox{\boldmath $\omega$}}

\newcommand{\bfpsi}{\mbox{\boldmath $\psi$}}

\newcommand{\bfPsi}{\mbox{\boldmath $\Psi$}}

\newcommand{\bfzero}{\mbox{\boldmath $0$}}
\newcommand{\bfone}{\mbox{\boldmath $1$}}

\newcommand{\diag}{\mathrm{diag}}
\newcommand{\blockdiag}{\mathrm{blockdiag}}

\newtheorem{algorithm}{Algorithm}[section]

\def\boxit#1{\vbox{\hrule\hbox{\vrule\kern6pt
          \vbox{\kern6pt#1\kern6pt}\kern6pt\vrule}\hrule}}
          
\begin{document}

\thispagestyle{empty} \baselineskip=28pt

\begin{center}
{\LARGE{\bf Spatiotemporal models for Poisson areal data with an application to the AIDS epidemic in Rio de Janeiro}}
\end{center}

\baselineskip=12pt

\vskip 2mm
\begin{center}
Marco A. R. Ferreira\footnote{(\baselineskip=10pt to whom
correspondence should be addressed) Department of Statistics,
Virginia Tech, 406-A Hutcheson Hall, 250 Drillfield Drive, Blacksburg, VA
24061, marf@vt.edu}, Juan C. Vivar\footnote{\baselineskip=10pt 
Food and Drug Administration}
\end{center}


\

\begin{center}
\section*{Abstract}
\end{center}
\baselineskip=12pt

We present a class of spatiotemporal models for Poisson areal data suitable for the
analysis of emerging infectious diseases. 
These models assume Poisson observations related through a link equation 
to a latent random field process. This latent random field process 
evolves through time with proper Gaussian Markov random field convolutions. 
Our approach naturally accommodates flexible structures such as distinct but interacting temporal 
trends for each region and across-time 
contamination among neighboring regions. 
We develop a Bayesian analysis approach with a simulation-based procedure:
specifically, we construct a Markov chain Monte Carlo algorithm based on the generalized extended Kalman filter 
to obtain samples from an approximate posterior distribution.  Finally, 
for the comparison of Poisson spatiotemporal models, we develop
a simulation-based conditional Bayes factor. 
We illustrate the utility and flexibility of our Poisson spatiotemporal framework 
with an application to the number of acquired immunodeficiency 
syndrome (AIDS) cases  during the period 1982-2007 in Rio de Janeiro.


\baselineskip=12pt
\par\vfill\noindent
{\bf Key Words:} Bayesian inference; Dynamic generalized linear models; Exponential family data; Regional data; State-space models.

\par\medskip\noindent

\clearpage\pagebreak\newpage \pagenumbering{arabic}
\baselineskip=24pt
\section{Introduction}
There is an increasing need for models for the spatiotemporal spread of emerging infectious diseases.
This increasing need arises from several factors such as the  impact of global warming on spread of insect vectors \citep{laff:2009,epst:2010} and higher global mobility
of goods and people \citep{apos:sonm:2007}. These factors have substantially  
increased the risk of epidemics of new, emerging, and re-emerging 
infectious diseases \citep{brow:chal:2003}.
In terms of publicly available information, 
datasets on diseases are usually in the form of areal count data; that is,
total number of cases of the disease available in a partition of the geographical 
domain of interest \citep{bane:carl:gelf:2014}. 
These areal count data are usually modeled using the Poisson distribution.
Several spatiotemporal models
for Poisson areal data have been developed in the disease mapping literature
\citep[e.g.,
][]{bern:1995,wall:carl:1997,knor:2000,schm:held:2004,tzal:best:2008}.
Usually, these models assume a common univariate temporal trend for all the regions
and time-specific spatial random effects. 
When modeling non-infectious diseases such as cancer, 
these models are usually effective
to explain the remaining spatiotemporal noise after covariates are taken into account. However,
these models are not appropriate for modeling the spatiotemporal dynamics of infectious diseases. 
In contrast, \citet{viva:ferr:2009} have introduced a general class of linear Gaussian
spatiotemporal models for areal data that allows for more complex spatiotemporal
processes.  
Even though the Gaussian model framework of \citet{viva:ferr:2009} may be applied to transformed count 
data when the counts are large, their framework is inadequate for dealing 
with counts that are small. Hence, their framework would not be appropriate for modeling the early epidemic expansion stages of an emerging disease.
Here we develop methodology that can be directly applied to areal count data.
Specifically, we propose a spatiotemporal class of models that allows for more complex spatiotemporal
processes for Poisson areal data.

There has been some previous literature where each subregion of the geographical 
domain of interest may have a different temporal trend
\citep{Sun:Tsut:Kim:He:2000,assu:reis:oliv:2001,macn:dean:2001}.
Typically, these models assume time-specific spatial random effects that follow a
Gaussian Markov random field. Furthermore, they assume region-specific temporal trends
that are deterministic functions of time and may be  linear \citep{Sun:Tsut:Kim:He:2000},
quadratic \citep{assu:reis:oliv:2001}, or piecewise cubic \citep{macn:dean:2001}. Finally,
\citet{Sun:Tsut:Kim:He:2000}, \citet{assu:reis:oliv:2001}, and \citet{macn:dean:2001} 
assume that the temporal trend coefficients follow Gaussian Markov random fields.
Even though these previous works allow for similar deterministic temporal trends for 
neighboring regions, they do not allow for more flexible temporal trends. 
Conversely, as we detail in Section~\ref{sec:models-for-epidemics},
our novel class of models allows for more flexible stochastic temporal trends
and for spatiotemporal interaction among the different regions.

There are two main modeling approaches to analyze epidemic data. One of these approaches 
is through compartmental
susceptible/exposed/infectious/ removed (SEIR) models \citep[e.g., see][]{ande:may:1992}. 
Based on the local behavior of individuals, these SEIR models assume
that the expected epidemic dynamics in the population 
can be represented by a set of partial differential equations. Because SEIR models are based
on the local behavior of individuals, \citet{mugg:cres:gemm:2002} say that SEIR models provide a ``particle''
view of the spread of an epidemic. 
The other modeling approach is what \citet{mugg:cres:gemm:2002}
call a ``field'' view. This field view approach pioneered by \citet{mugg:cres:gemm:2002} is particularly
 adequate 
for spatiotemporal areal data and models the stochastic geographic spread of the epidemic through 
time using a multivariate dynamic generalized linear model. Specifically, 
\citet{mugg:cres:gemm:2002} model the spatiotemporal dynamics of an influenza epidemic 
with what they call a spatially descriptive temporally dynamic hierarchical model.
Their model is a particular case of our class of models where they use what we
call a contamination model.
Similarly to \citet{mugg:cres:gemm:2002}, here we take a field view 
of the spatiotemporal epidemic process. 

To perform Bayesian statistical analysis for our proposed class of spatiotemporal 
models, here we develop a simulation-based procedure. 
Specifically, we construct a Markov chain Monte Carlo (MCMC)
algorithm \citep{gelf:smit:1990,robe:case:2004,game:lope:2006} to obtain samples from
an approximate posterior distribution. The MCMC algorithm we propose 
combines the generalized extended Kalman filter \citep[p. 352,][]{fahr:tutz:2001} and the forward filter
backward sampler method (FFBS) \citep{fruh:1994,cart:kohn:1994} for the simulation of the latent 
fields. Our computational methodology is general enough to be applied to observations
with any distribution in the regular exponential family. However, here we focus on Poisson observations.

With the ability to fit several nonnested spatiotemporal models comes the need to
compare those models. In addition, 
when analyzing a particular dataset, each of the fitted models corresponds to a
distinct scientific hypothesis. As such, comparison of the different models becomes
paramount for understanding the nature of the process underlying the observed data.
For the comparison of the resulting nonnested Poisson spatiotemporal models, we develop
a conditional Bayes factor \citep[p. 190,][]{ghos:dela:sama:2006} computed  
by using the output of our MCMC algorithm. As we discuss in Section~\ref{sec:model-selection},
this conditional Bayes factor provides a model comparison criterion
that is justified from a Bayesian decision-theoretic point of view \citep{berg:1985}.
In addition,
this conditional Bayes factor is computed based on the one-step ahead predictive
densities of the competing models. 
Therefore, as an important practical consequence, this conditional Bayes factor 
favors models that provide better probabilistic predictions.

We illustrate the utility and flexibility of our Poisson spatiotemporal framework 
with an application to the number of acquired immunodeficiency 
syndrome (AIDS) cases in Rio de Janeiro State, Brazil.
Specifically, 
we consider the annual number of cases of AIDS from 1982 to 2007 for each of the
92 counties in Rio de Janeiro State.
These data are publicly available from the Ministry of Health of Brazil
and may be downloaded from the webpage https://datasus.saude.gov.br.
AIDS is caused by the immunodeficience virus (HIV), which has a long incubation 
period with a median of about 10 years in young adults \citep{bacc:moss:1989}.
Because the incubation period is much longer than the observational time unit 
of one year, we may expect the observed number of cases in one year to be
a convolution of infections that have occurred in several previous years. 
As a result, the change from year to year in the underlying latent expected number of cases
may have a strong spatial dependence.
Finally, given the time scale of the incubation period of about one decade
and the fact that AIDS has emerged about four decades ago, the AIDS epidemic 
provides an excellent case study of what may be the spatiotemporal dynamics 
when a disease is emerging in a previously unaffected region.

This chapter is organized as follows: Section~\ref{secmodel}
presents the proposed general class of spatiotemporal Poisson models for
epidemic areal data, as well as several useful specific models.
Section~\ref{sec:bayinf} develops simulation-based Bayesian estimation 
and model selection for these spatiotemporal models.
Section~\ref{sec:applic} illustrates our spatiotemporal framework
with an application to the number of AIDS cases in the State of Rio
de Janeiro, Brazil. Conclusions and possible extensions are
presented in Section~\ref{sec:conclusions}.

\section{Spatiotemporal Models for Poisson Areal Data}\label{secmodel}

Consider a geographical domain of interest partitioned in a collection of $S$ regions
indexed by the integers $1, \ldots, S$ forming a regular or irregular grid.  
Assume that this grid is endowed with a
neighborhood system $\{N_s, s=1,\ldots,S\}$, where $N_s$ is the set of regions
that are neighbors of region $s$. Let $y_{ts}$ be the number of cases observed
at time $t$ on region $s$, $t = 1, \ldots, T, \ s = 1, \ldots, S$.
As usual for count data, assume that the observation
$y_{ts}$ follows a Poisson distribution. Specifically, assume
that $y_{ts}|\lambda_{ts} \sim Po(n_{ts}\lambda_{ts})$, where
$n_{ts}$ denotes the population size and $\lambda_{ts}$ is the
underlying risk at time $t$ and region $s$, $t = 1, \ldots, T,
s = 1, \ldots, S$. Thus, the probability function of $y_{ts}$ conditional on
$\lambda_{ts}$ is
\begin{equation}
p(y_{ts}|\lambda_{ts}) =   \frac{(n_{ts} \lambda_{ts})^{y_{ts}} e^{-n_{ts} \lambda_{ts}}}{y_{ts}!}.   
\label{eqn:observational}
\end{equation}
Let $\eta_{ts} = \log(n_{ts} \lambda_{ts})$, $a=1$, $b(\eta_{ts}) = n_{ts} \lambda_{ts} = \exp(\eta_{ts})$, 
and $c(y_{ts}) = -\log(y_{ts}!)$. Then, the Poisson probability function can be rewritten as 
\citep{case:berg:2001}
\begin{eqnarray} \label{obs_eq}
p(y_{ts} | \eta_{ts}, \upsilon) & \propto & \exp\{(y_{ts} \eta_{ts}
- b(\eta_{ts}))/a + c(y_{ts})\},
\end{eqnarray}
and thus belongs to the regular
exponential family of distributions. Here, the natural parameter is $\eta_{ts} =
\log(\mu_{ts})$, where $\mu_{ts}=b'(\eta_{ts})=exp(\eta_{ts})=n_{ts}\lambda_{ts}$ is 
the mean of $y_{ts}$. Hence, we may rewrite the natural parameter as
$\eta_{ts} = \log n_{ts} + \log \lambda_{ts}$.


%

To model the mean level,  
generalized linear models assume a monotone
differentiable link function $g(.)$ that transforms the
mean $\mu_{ts}$ from its restricted range to the unrestricted real line \citep{mccu:neld:1996}. 
For example, in the case of Poisson data a typical choice 
is the canonical link function $g(\mu)=\log \mu$. In the spatiotemporal case we consider, 
the canonical link function becomes
$g(\mu_{ts}) = \log(\mu_{ts}) = \log n_{ts} + \log \lambda_{ts}$,
where $\log n_{ts}$ is a known offset. In this case, we define $\theta_{ts} = \log \lambda_{ts}$
as the value of the latent log-risk random field at time $t$ and region $s$.
Then, we may directly relate $\theta_{ts}$ to the natural parameter $\eta_{ts}$
through the function $\delta_{ts}(.)$ defined as
\begin{eqnarray} 
\eta_{ts}   & = & \delta_{ts}(\theta_{ts}) \ = \ \theta_{ts}+ \log n_{ts}. \label{lat_eq1} 
\end{eqnarray}
Finally, assuming the canonical link function $g(\mu_{ts}) = \log(\mu_{ts}) = \log n_{ts} + \log \lambda_{ts}$
implies that the so-called response function, i.e. the inverse of the link function,
is $f(\theta_{ts}) = n_{ts} \exp(\theta_{ts})$. The response function plays an important role in the construction
of the generalized extended forward filter detailed in Algorithm 3.1.

Let $\bftheta_t = (\theta_{t1}, \ldots, \theta_{tS})'$ be the vectorized latent 
log-risk random field at time $t$. We relate $\bftheta_t$ to a dynamic latent vector
$\bfbeta_t$ through the linear equation
\begin{eqnarray} 
\bftheta_t  & = & \bfF'_t \bfbeta_t, \hspace{2.2cm}  \label{lat_eq2}
\end{eqnarray}
where $\bfF_t$ is typically known up to some few unknown hyperparameters.
Finally, we assume that the latent vector $\bfbeta_t$ evolves through time 
following the system (also known as evolution or state-space) equation
\begin{eqnarray} 
\bfbeta_t   & = & \bfG_t \bfbeta_{t-1} + \bfomega_t, \hspace{1cm}
\bfomega_t \sim PGMRF(\bfzero, \bfW_t^{-1}), \label{lat_eq3}
\end{eqnarray}
where the evolution matrix $\bfG_t$ is usually known up to some few unknown
hyperparameters. Equation~(\ref{lat_eq3}) generalizes the dynamic generalized linear model system equation
of \citet{west:harr:migo:1985,prad:ferr:west:2021} to the multivariate 
spatiotemporal case.
Further, 
$\bfomega_t$ is a vector of evolution innovations that follows the proper Gaussian 
Markov random field (PGMRF)
process discussed in Section~\ref{sec:ferreira-deoliveira-mrfs}
with zero mean vector and precision matrix $\bfW_t^{-1}$, that is,
the density of $\bfomega_t$ is proportional to
$$
|\bfW_t|^{-0.5} \exp(-0.5 \bfomega_t' \bfW_t^{-1} \bfomega_t),
$$
where the precision matrix is $\bfW_t^{-1}
= \tau(\bfI + \phi \bfM)$. Here, $\tau>0$ is a scale parameter,
$\phi>0$ controls the amount of spatial correlation, and $\bfI$ is the 
identity matrix. In addition, the neighborhood matrix $\bfM$ 
is such that
$(\bfM)_{k,l} = m_{k} \mbox{ if } k=l,$  $(\bfM)_{k,l} =  -g_{kl} \mbox{ if }  \
k \in N_l,$
and $(\bfM)_{k,l} = 0 \mbox{ otherwise},$
where
$g_{kl} > 0$
is a measure of similarity between regions $k$ and $l$, and the diagonal elements are defined as $m_k =                  
\sum_{l \in N_k} g_{kl}, k=1,\ldots,S$.
For additional details on this type of PGMRFs, see \citet{ferr:oliv:2007}.

\subsection{Spatiotemporal Models for Epidemics} \label{sec:models-for-epidemics}

This section discusses several important specific models within the general
class of models defined by Equations (\ref{eqn:observational}), 
(\ref{lat_eq1}), (\ref{lat_eq2}), and (\ref{lat_eq3}) that are useful for spatiotemporal 
modeling of epidemics. 
Epidemic processes of emerging diseases, as is the case of the AIDS
data that we consider, are usually nonstationary at the start of the epidemic.
Thus, here we consider five distinct nonstationary state-space equation specifications. 
Further, for each of these five specifications, we consider two distinct state-space covariance 
matrices, comprising a total of ten models. The first covariance matrix is diagonal, and the
second covariance matrix is defined (as discussed above) as the covariance matrix of a PGMRF.

The first state-space equation specification we consider is a first-order temporal trend evolution.
Specifically, the first-order temporal trend evolution assumes 
${\bfF}_t' = {\bfI}_S$ and ${\bfG}_t = {\bfI}_S$. This matrix ${\bfF}_t$ implies that $\beta_{ts}$
is the log of the risk  at time $t$ for each person in subregion~$s$.
When the state-space precision matrix is the PGMRF matrix
${\bfW}_t^{-1} = \tau({\bfI}_S + \phi{\bfM})$,
the implied expected value of $\beta_{ts}$, $t=2,\ldots, T-1$, conditional on the entire latent process
is $0.5(\beta_{t-1,s}+\beta_{t+1,s})+m_s^{-1}\sum_{k \in N_s}\{\beta_{tk}-0.5(\beta_{t-1,k}+\beta_{t+1,k})\}$.
Hence, the conditional mean of $\beta_{ts}$ depends on the latent process in region $s$ at times $t-1$ and $t+1$,
as well as the latent process in the spatial neighbors of region $s$ at times $t-1$, $t$, and $t+1$. Thus, the 
first-order temporal trend evolution may be used for spatiotemporal smoothing.
Therefore, this state-space specification may be useful after the epidemic ends its expansion
phase and the disease becomes endemic.

The second state-space specification incorporates spatiotemporal contamination.
Specifically, evolution with contamination assumes 
${\bfF}_t' = {\bfI}_S$, and a contamination matrix ${\bfG}_t = (1 + \kappa h)^{-1} \bfH$,
with $\{{\bfH}\}_{kl} = 1$ for $k=l$, $\{{\bfH}\}_{kl} = \kappa$ for $k \in C_l$, and $\{{\bfH}\}_{kl} = 0$
otherwise. Here, $C_l$ is the set of across-time neighbors of subregion $l$, 
$h = \max_k \bfone(k \in C_l)$ is the maximum number of across-time neighbors, 
and $\kappa>0$ is a contamination coefficient. Note that for full generality the
set of across-time neighbors  $C_l$ may differ from the set of time-specific neighbors $N_l$.
This contamination specification allows for abnormal increases at a given
time point to spill over to neighboring regions at subsequent time points.  Thus, we expect this
contamination state-space specification to be useful during the initial periods of a disease emerging in a previously unaffected region.

The third  state-space specification assumes a second-order temporal trend.
The second-order temporal trend model assumes that $\bfbeta_t$ is comprised of
two latent fields at time $t$, $\bfbeta_{1t}=(\beta_{1 t 1},\ldots,\beta_{1 t S})'$ and 
$\bfbeta_{2t}=(\beta_{2 t 1},\ldots,\beta_{2 t S})'$. Hence, this model assumes
$\bfbeta_t = (\bfbeta_{1t}',\bfbeta_{2t}')'$. In this specification, 
$\bfbeta_{1t}$ is equal to $\bftheta_t$, the log-risk random field at time $t$.
Furthermore, $\bfbeta_{2t}$ is the gradient field, that is, the expected increase in
the log-risk random field from time $t$ to time $t+1$.
Specifically, the second-order temporal trend model assumes
${\bfF}'_t = ({\bfI}_S, {\bfzero}_{SS})$, and evolution matrix
$${\bfG}_t = \left(
\begin{array}{cc}
     {\bfG}_{1t} & {\bfG}_{1t} \\
           {\bfzero}_{SS} & {\bfG}_{2t}
\end{array} \right),
$$
where ${\bfG}_{it} = {\bfI}_S$, $i = 1, 2$, and ${\bfzero}_{SS}$ is an $S \times S$ matrix
of zeros.
Further, in the case of spatially correlated state-space innovations, we assume
that ${\bfW}_t^{-1}$ is block diagonal, that is, 
${\bfW}_t^{-1} = \blockdiag({\bfW}_{1t}^{-1}, {\bfW}_{2t}^{-1})$,
where
${\bfW}_{it}^{-1} = \tau_i({\bfI}_S + \phi_i{\bfM})$, $i = 1, 2.$ The 
second-order temporal trend model is appropriate when the temporal trend
for each region can be approximated by a local linear trend that changes
smoothly over time. Therefore, this model may be useful for describing the 
spatiotemporal evolution of the early stages of an emerging disease.

The fourth state-space specification assumes a field log-risk
level and a common gradient for all counties at each time point.
The common gradient at time $t$ is a univariate parameter $\beta_{2t}$ and
evolves through time accordingly to the evolution equation $\beta_{2t} = \beta_{2,t-1}+\omega_{2t}$,
where $\omega_{2t} \sim N(0,\psi^{-1})$ and $\psi$ is the evolution precision for the common
gradient. For this state-space specification, $\bfbeta_t = (\bfbeta_{1t}',\beta_{2t})'$, ${\bfF}'_t = ({\bfI}_S, {\bfzero}_S)$, and the evolution matrix is
$${\bfG}_t = \left(
\begin{array}{cc}
     {\bfI}_{S} & {\bfone}_{S} \\
           {\bfzero}_S' & 1
\end{array} \right),
$$
where ${\bfzero}_S$ and ${\bfone}_S$ are $S$-dimensional vectors of zeros and ones, respectively.
For the fourth state-space specification, the state-space precision matrix is
$${\bfW}_t^{-1} = \left(
\begin{array}{cc}
     {\bfW}_{1t}^{-1} & {\bfzero}_{S} \\
           {\bfzero}_S' & \psi^{-1}
\end{array} \right).
$$

Finally, the fifth state-space specification assumes a contamination
field model for the log-risk combined with a common gradient for all counties at each time point. 
Hence, almost all the components of the fifth state-space specification are the same 
as the components of the fourth state-space specification, except that the evolution matrix
includes a contamination matrix and is given by
$${\bfG}_t = \left(
\begin{array}{cc}
    (1 + \kappa h)^{-1} \bfH  & {\bfone}_{S} \\
           {\bfzero}_S' & 1
\end{array} \right),
$$
where $\kappa$, $h$, and $\bfH$ are defined as in the second state-space specification
described above.

The models we describe above by no means exhaust the general class of models
defined by Equations  (\ref{eqn:observational}), (\ref{lat_eq1}), (\ref{lat_eq2}), and (\ref{lat_eq3}).
In particular, many other models may be obtained as stochastic discretized versions of
deterministic continuous time-space mathematical models. For example, these deterministic continuous 
time-space mathematical models may be based on partial differential equations
and on integro-difference equations. For details on several possible deterministic continuous 
time-space mathematical models and how to discretize them, see \citet{wikl:hoot:2010} and 
references therein.

\section{Bayesian Inference} \label{sec:bayinf}

This section develops simulation-based methods to perform full Bayesian statistical
analysis for the class of spatiotemporal models defined by Equations 
(\ref{eqn:observational}), (\ref{lat_eq1}), (\ref{lat_eq2}), and (\ref{lat_eq3}). 
Specifically, we construct a Markov chain Monte Carlo (MCMC)
algorithm \citep{gelf:smit:1990,robe:case:2004,game:lope:2006} to obtain samples from
an approximate posterior distribution.
Our computational methodology is general enough to be applied to observations
with any distribution in the regular exponential family. However, here we focus on Poisson observations.

Let $\bfbeta_{0:T} = (\bfbeta_0',\ldots,\bfbeta_T')'$ be the vectorized latent process collected
through time, with similar notation for other quantities of the model. 
In addition, let $\bfpsi$ be the vector of unknown hyperparameters.
Further, let $\bfD_t$, $t=1, \ldots, T$, represent all the information up to time $t$. 
Thus, $\bfD_t$ is recursively defined as $\bfD_t=\bfD_{t-1}\cup \{\bfy_t\}$.
Then, by Bayes Theorem the exact posterior density for $(\bfbeta_{0:T}, \bfpsi)$ is proportional to
\begin{eqnarray*} 
p(\bfbeta_{0:T}, \bfpsi | \bfD_T) & \propto&
\left\{\prod_{t=1}^{T}{p\left(\bfy_t|\bftheta_t,\bfeta_t\right)}\right\}
\left\{\prod_{t=1}^{T}{p\left(\bfbeta_t|\bfbeta_{t-1}, \bfpsi\right)}\right\}
p\left(\bfbeta_0 | \bfpsi\right)
p(\bfpsi).
\end{eqnarray*}

The Markov chain used in our procedure has to be tailored to each specific 
spatiotemporal model and will depend on how the hyperparameter vector $\bfpsi$ 
appears in the matrices ${\bfF}_t, {\bfG}_t$ and
${\bfW}_t$. However, the Markov chain may
be partitioned in two blocks: simulation of ${\bfpsi}$, which is model specific, and
simulation of the latent process $({\bfbeta}_0,\ldots,{\bfbeta}_T)$. For the simulation of the latent
process, we propose in Section~\ref{sec:EFFBS} a novel extended forward filter backward sampler (EFFBS).

\subsection{Extended Forward Filter Backward Sampler} \label{sec:EFFBS}

This section introduces a novel extended forward filter backward sampler for the 
simulation of the latent process $({\bfbeta}_0,\ldots,{\bfbeta}_T)$. The
EFFBS we propose combines the generalized extended Kalman filter  
\citep[p. 352,][]{fahr:tutz:2001} and the forward filter
backward sampler method (FFBS) \citep{fruh:1994,cart:kohn:1994}.
Hence, the EFFBS is composed of two subsequent steps, that we describe
below as the extended forward filter in Algorithm 3.1 and the
backward sampler in Algorithm 3.2.

Let 
$\widehat{\bftheta}_t =
(\widehat{\theta}_{t1}, \ldots, \widehat{\theta}_{tS})'$
be a preliminary estimate of $\bftheta_t$. This preliminary estimate may, for example, be 
the posterior mode obtained
by the generalized extended Kalman filter. 
In addition, let $f'({\theta}_{ts})=n_{ts} e^{\theta_{ts}}$ denote the first derivative of the 
response function $f({\theta}_{ts})$.
Furthermore, let $\Sigma_{ts}= Var(y_{ts})=n_{ts} e^{\theta_{ts}}$ be the variance of the
number of disease cases at time $t$ and region $s$.
Then, the extended forward filter proceeds as follows.

\begin{algorithm}[Extended forward filter] Initiate the algorithm at time $t=0$ with the 
distribution $\bfbeta_0|\bfpsi, \bfD_0 \sim N\left({\bfm}_0,{\bfC}_0\right)$. Then, for $t=1, \ldots, T$, do:

\begin{enumerate}
\item Prior at $t$: $\bfbeta_t|\bfpsi, \bfD_{t-1} \sim
N(\bfa_t,\bfR_t)$, with \begin{eqnarray}
\bfa_t &=& \bfG_t \bfm_{t-1}, \nonumber \\
\bfR_t &=& \bfG_t \bfC_{t-1} \bfG_t' + \bfW_t. \nonumber
\end{eqnarray}

\item Based on the first-order Taylor expansion of the response function $f$ around
$\widehat{\bftheta}_t = \bfF'_t \bfa_t$, compute the
artificial observation vector  $\widehat{\bfy}_t =
(\widehat{y}_{t1}, \ldots,$ $\widehat{y}_{tS})$ with
\begin{eqnarray} \label{eqn:artificial-observation}
\widehat{y}_{ts} & = &
\left\{f'(\widehat{\theta}_{ts})\right\}^{-1}\left\{y_{ts} -
f(\widehat{\theta}_{ts})\right\} + \widehat{\theta}_{ts}, 
\end{eqnarray}
and the approximate precision matrix
\begin{eqnarray} \label{eqn:approx-prec-matrix}
\widehat{\bfV}_t^{-1} & = &
\diag\left[\frac{\left\{f'\left(\widehat{\theta}_{ts}\right)\right\}^2}{\widehat{\Sigma}_{ts}}\right].
\end{eqnarray}

\item Posterior at $t$: $\bfbeta_t|\bfpsi, \bfD_t
\sim N\left({\bfm}_t,{\bfC}_t\right)$, with
\begin{eqnarray}
{\bfC}_t & = & \left(\bfR_t^{-1} + \bfF_t \widehat{\bfV}_t^{-1} \bfF_t'\right)^{-1}, \\
{\bfm}_t & = & {\bfC}_t \left(\bfF_t \widehat{\bfV}_t^{-1}
\widehat{\bfy}_t + \bfR_t^{-1} \bfa_t\right).
\end{eqnarray}
\end{enumerate}

\end{algorithm}

The extended forward filter is an approximate method to calculate the
mean vectors and covariance matrices of the posterior distributions
$p(\bfbeta_t|\bfpsi, \bfD_t)$, $t=1,\ldots,T$. At the end of the extended forward filter, we have an approximation
for the posterior distribution of the latent process at the last time point $T$ given all the available 
information as well as the hyperparameter vector $\bfpsi$, that is, $p(\bfbeta_T|\bfpsi, \bfD_T)$. Then, the backward sampler proceeds as follows.

\begin{algorithm}[Backward sampler] Sampling from $\bfbeta_{1:T} | \bfpsi, \bfD_t$: 

\begin{enumerate}
\item Sample $\bfbeta_T^*$  from the distribution $N({\bfm}_T, {\bfC}_T)$, 
obtained from the extended forward filter. 
\item
For $t=T-1,\ldots,0$, sample backwards $\bfbeta_{t}^*$  from the conditional
distribution $N({\bfb}_t, {\bfB}_t)$, where
\begin{eqnarray}
\label{eqn:effbs1} {\bfB}_t & = & \left({\bfC}_t^{-1} +
\bfG_{t+1}'\bfW_{t+1}^{-1}\bfG_{t+1}\right)^{-1}, \\
\label{eqn:effbs2} {\bfb}_t & = & {\bfB}_t
\left(\bfG_{t+1}'\bfW_{t+1}^{-1}\bfbeta_{t+1}^* +
{\bfC}_t^{-1}{\bfm}_t\right).
\end{eqnarray}

\end{enumerate}

\end{algorithm}

The EFFBS assumes an approximate Gaussian observational equation with artificial observation vector defined in (\ref{eqn:artificial-observation}) and approximate precision matrix
given by (\ref{eqn:approx-prec-matrix}). This implies that the EFFBS implicitly assumes 
the approximate likelihood function
\begin{equation} \label{eqn:approx_likelihood}
{k\left(\bfy_{1:T}|\bftheta_{1:T},\bfpsi\right)} = \prod_{t=1}^{T}{k\left(\bfy_t|\bftheta_t,\bfpsi\right)}, 
\end{equation}
where $k\left(\bfy_t|\bftheta_t,\bfpsi\right)$ is proportional to
\begin{align*}
\left( \prod_{s=1}^S  \widehat\Sigma_{ts} \right)^{-1/2}
\exp\Bigg[  -0.5 \sum_{s=1}^S (n_{ts} e^{\widehat\theta_{ts}})^{-1} \bigg\{ & y_{ts}-n_{ts}e^{\widehat\theta_{ts}}  \\
& + n_{ts} e^{\widehat\theta_{ts}} \left(\widehat\theta_{ts} - (\bfF_t' \bfbeta_t)_s\right)
\bigg\}^2 \Bigg].
\end{align*}
Therefore, the MCMC algorithm with the embedded EFFBS defines a Markov chain that converges 
to an approximate posterior distribution with density
\begin{eqnarray*} 
k(\bfbeta_{0:T}, \bfpsi | \bfD_T) & \propto&
\left\{\prod_{t=1}^{T}{k\left(\bfy_t|\bftheta_t,\bfpsi\right)}\right\}
\left\{\prod_{t=1}^{T}{p\left(\bfbeta_t|\bfbeta_{t-1}, \bfpsi\right)}\right\}
p\left(\bfbeta_0 | \bfpsi\right)
p(\bfpsi).
\end{eqnarray*}

Finally, at the end of the MCMC algorithm we have a sample 
$\left(\bfpsi^{(1)},\bfbeta_{0:T}^{(1)}\right), \ldots,$ 
$\left(\bfpsi^{(G)},\bfbeta_{0:T}^{(G)}\right)$ 
from this approximate posterior distribution.

\subsection{Model Comparison} \label{sec:model-selection}

With the ability to fit several nonnested spatiotemporal models comes the need to
compare those models. Note also that, 
when analyzing a particular dataset, each of the possible models corresponds to a
distinct scientific hypothesis. As such, comparison of the different models becomes
paramount for understanding the nature of the process underlying the observed data.
For the comparison of the nonnested spatiotemporal models introduced in 
Section~\ref{secmodel}, we develop
a conditional Bayes factor \citep[p. 190,][]{ghos:dela:sama:2006} that uses the output of the MCMC algorithm for the approximate posterior
distribution. As we discuss below,
this conditional Bayes factor provides a model comparison criterion
that is justified from a Bayesian decision-theoretic point of view \citep{berg:1985}.

Bayesian model selection is usually performed by comparing the
posterior probabilities of the competing models. When the competing
models have equal prior probabilities, their posterior probabilities
are proportional to their respective predictive densities. A potential
difficulty in Bayesian model selection is that these predictive 
densities will be sensitive to the specification of the prior distribution for the hyperparameter
vector ${\bfpsi}$. To overcome this difficulty, we use here
a training sample approach \citep{fruh:1995} of the first $t^*$ time
observations; this results in calibrated priors for the parameters
of each model. Then, Monte Carlo integration is used to compute the
predictive distribution under each model for the remaining $T-t^*$
time observations.

Suppose that there are $Q$ competing spatiotemporal models denoted by
$M_1,\ldots,M_Q$. The $q$th model has observational density
$p_q({\bfy}_t| {\bfbeta}_t,\bfPsi)$ and evolution density
$p_q({\bfbeta}_t|{\bfbeta}_{t-1}, {\bfPsi})$.
Let $p_q({\bfbeta}_{0:(t-1)},{\bfPsi}|{\bfD}_{t-1})$ denote the joint
posterior distribution of ${\bfPsi}$ and ${\bf
\beta}_0,\ldots,{\bfbeta}_{t-1}$ under the $q$th model given the 
information up to time $t-1$.
Note that the definitions of the hyperparameter vector ${\bfPsi}$ and the
latent vector ${\bfbeta}_t$ may be (and in general will be)
different under each model, but this distinction is omitted here
to keep the notation simple.
Then, under model $q$ the one-step ahead predictive density of ${\bfy}_t$ 
given the information up to time $t-1$ is 
\begin{align}
p_q(\bfy_t|\bfD_{t-1}) = \int &p_q(\bfy_t|\bfbeta_t,
\bfPsi) p_q(\bfbeta_t|\bfbeta_{t-1},
\bfPsi) \nonumber \\
& \times 
p_q(\bfbeta_{0:(t-1)},
\bfPsi|\bfD_{t-1}) d\bfbeta_{0:(t-1)}
d\bfbeta_t d\bfPsi. \label{eq1}
\end{align}

We apply the simulation scheme outlined in
Section~\ref{sec:bayinf} to obtain $G$ posterior draws 
$$\left[\left(\bfbeta^{(1)}_{0:(t-1)}, \bfPsi^{(1)}\right), \left(\bfbeta^{(2)}_{0:(t-1)},
\bfPsi^{(2)}\right), \ldots, \left(\bfbeta^{(G)}_{0:(t-1)},
\bfPsi^{(G)}\right)\right].$$
Next, for  $g=1,\ldots,G$, we simulate $\bfbeta_t^{(g)}$ given
$(\bfbeta^{(g)}_{0:(t-1)}, \bfPsi^{(g)})$ from the evolution distribution with density
$p_q(\bfbeta_t|\bfbeta^{(g)}_{0:(t-1)}, \bfPsi^{(g)})$. 
This yields the posterior sample $[(\bfbeta^{(1)}_{0:t}, \bfPsi^{(1)}), \ldots,
(\bfbeta^{(G)}_{0:t}, \bfPsi^{(G)})]$.

Now rewrite Equation (\ref{eq1}) as
\begin{equation}\label{eq2}
p_q(\bfy_t|\bfD_{t-1}) = \int{p_q(\bfy_t|\bfbeta_t, \bfPsi)
p_q(\bfbeta_{0:t}, \bfPsi|\bfD_{t-1}) d\bfbeta_{0:t} d\bfPsi}.
\end{equation}
Equation (\ref{eq2}) shows that $p_q(\bfy_t|\bfD_{t-1})$ is the expectation 
of $p_q(\bfy_t|\bfbeta_t, \bfPsi)$ with
respect to $p_q(\bfbeta_{0:t}, \bfPsi | \bfD_{t-1})$. Therefore, we can 
estimate the one-step ahead predictive density $p_q(\bfy_t|\bfD_{t-1})$ with
\begin{equation}
\widehat{p}_q(\bfy_t|\bfD_{t-1}) = \frac{1}{G}\sum_{g=1}^{G}
p_q(\bfy_t|\bfbeta_t^{(g)}, \bfPsi^{(g)}). \label{eqn:one_step_predictive}
\end{equation}

Using the fact that the
joint predictive density of ${\bfy}_{t^*+1}, \ldots,
{\bfy}_T$ can be written as $p_q({\bfy}_{t^*+1},
\ldots, {\bfy}_T|{\bfD}_{t^*}) = \prod_{t=t^*+1}^{T}{p_q({\bf
y}_t|{\bfD}_{t-1})}$, 
and as $\bfD_t=\bfD_{t-1}\cup \{\bfy_t\}$,
an estimate of the joint predictive density
under model $q$ is \citep{viva:ferr:2009}
\begin{eqnarray} \label{eqn:joint_predictive}
\widehat{p}_q({\bfy}_{t^*+1}, \ldots, {\bfy}_T|{
\bfD}_{t^*}) & = & \prod_{t=t^*+1}^{T}{\widehat{p}_q({\bfy}_t|{\bfD}_{t-1})}, 
\end{eqnarray}
where $t^*$ is such that $p_q({\bfpsi}|{\bfD}_{t^*})$ is proper for
all $q=1,\ldots,Q$. For a detailed discussion on the choice of $t^*$, see \citet{viva:ferr:2009}.
Then, based on the joint predictive density~(\ref{eqn:joint_predictive}), 
for each pair of models we compute the conditional Bayes factor of model $m$ against model~$n$~as
$$                                                                                                                                                    
B_{m n} = \frac{\hat{p}_m(\bfy_{t^*+1}, \ldots, \bfy_T|\bfD_{t^*})}{\hat{p}_n(\bfy_{t^*+1}, \ldots, \bfy_T|\bfD_{t^*}) }.                             
$$
Finally, we use the conditional Bayes factors to decide what is the best model.
In addition to being well justified theoretically, these conditional Bayes factors 
favor models that provide better probabilistic predictions.

\section{Application}\label{sec:applic}

We illustrate the utility and flexibility of our Poisson spatiotemporal framework 
with an application to the number of acquired immunodeficiency 
syndrome (AIDS) cases in Rio de Janeiro State, Brazil.
Specifically, 
we consider the annual number of new cases of AIDS from 1982 to 2007 for each of the
92 counties in Rio de Janeiro State. 
Data on the annual number of new cases of AIDS and on population size are publicly available from the Ministry of Health of Brazil
and may be downloaded from the webpage www.datasus.saude.gov.br.
From 1982 to 2007 new counties were created in Rio de Janeiro
State. The data available on www.datasus.saude.gov.br about the number of new cases already
use the geopolitical organization of Rio de Janeiro State as of 2007. However, data on population
sizes are available for each year based on the geopolitical organization for each specific year.
We have performed backcasting to obtain estimates of annual population sizes per county using the 
geopolitical organization of Rio de Janeiro State as of 2007. The analysis that we present below is
conditional on these county population sizes estimates.

AIDS is caused by the immunodeficience virus (HIV), which has a long incubation 
period with a median of about 10 years in young adults \citep{bacc:moss:1989}.
Because the incubation period is much longer than the observational time unit 
of one year, we may expect the observed number of cases in one year to be
a convolution of infections that have occurred in several previous years. 
As a result, the change from year to year in the underlying risk of observing new cases
may exhibit a strong spatial dependence.
This spatial dependence is accounted for in our modeling framework through the system
equation innovation $\bfomega_t$ and its covariance matrix $\bfW_t$.
Finally, given the time scale of the incubation period of about one decade
and the fact that AIDS has emerged among humans about four decades ago, the AIDS epidemic 
provides an excellent case study of what may happen in terms of spatiotemporal dynamics 
when a previously unaffected region becomes infected by an emerging infectious disease.

We have implemented the Bayesian analysis procedures developed in Section~\ref{sec:bayinf}
for each of the five spatiotemporal models described in Section~\ref{sec:models-for-epidemics}.
Specifically, we
consider the following models with spatially dependent innovations: I -- First-order field polynomial model;
II -- Contamination field model; III - Second-order field polynomial model; IV - Model with a field log-risk
level and a common gradient for all counties at each time point; V - Contamination
field model for the log-risk combined with a common gradient for all counties at each time point. 
Further, we have considered for each of these five models two specifications for the system
equation innovations: spatial dependence, and spatial independence. 

A Bayesian analysis requires the specification of priors both for the hyperparameters
and for the latent process $\bfbeta_0$ at time $t=0$ given  the information up to
time $0$ denoted by $\bfD_0$. For the specification of the prior for $\bfbeta_0$ for an emerging disease,
we may use the information about the type of disease that we anticipate to observe.
In the case of slowly progressing emerging epidemics such as in the case of AIDS in the beginning of the 1980s, we expect the standardized
risk for each county of the first
case to appear at a given time to be about the reciprocal of the state population size.
In the case of Rio de Janeiro State, the population size in 1981 (t=0) was about 11.5 million people.
Hence, this corresponds to an expected logarithm of the risk equal to 
$E(\bfF' \bfbeta_0 | \bfD_0)= \bfone_S \; log(1/11,500,000) \approx -16.25 \;  \bfone_S$. In addition, 
for an emerging disease we believe that
the risk of the first case to appear at a given time to be within 20-fold of its expectation with
probability 0.997. Further, we assume that before the disease emerges the standardized risk 
is stochastically independent between counties. These assumptions correspond 
in the logarithm scale to assuming $Var(\bfF' \bfbeta_0 | \bfD_0)= \bfI_S$.

When assigning priors for the hyperparameters, we follow closely the recommendations 
of \citet{viva:ferr:2009}. Specifically, we assume for the scale parameter $\tau_i$ a weakly informative
$Gamma(1,1)$ prior; this prior imparts little information and concurrently guarantees
posterior propriety. In addition, for the contamination coefficient 
$\kappa$ we assume a noninformative uniform prior on the interval $(0,1)$.
Further, for the spatial dependence parameter $\phi_i$ we assign a prior of the type
$\Pi(\phi_i) \propto 1 \mbox{ if } 0 < \phi_i <1, $
$\Pi(\phi_i) \propto  \phi_i^{-2}  \mbox{ if } 1 < \phi_i <a,$
and $\Pi(\phi_i) \propto  0$ otherwise. 
See \citet{viva:ferr:2009} for an explanation on the relationship of this prior and the
marginal reference prior for $\phi_i$ proposed by \citet{ferr:oliv:2007} for PGMRFs.
In particular, we have found that $a=100$ works well
in our current application. 

Finally, for Models 4 and 5 that include a common gradient we assign a conditionally conjugate
prior $Gamma(a_\psi,b_\psi)$ for the log-gradient evolution precision $\psi$.
First, to obtain a reasonably vague prior we assume $b_\psi=0.1$ that implies the
prior variance for $\psi$ to be 10 times larger than the prior mean. Moreover,
we expect the common gradient of the risk to be reasonably stable
at subsequent time points. In particular, we expect the gradient to do not vary by much more than 
20\% between subsequent time points. This informs us that in the logarithm scale
the evolution standard deviation $\psi^{-1/2}$ 
is probably less than 0.1; hence, we assign for this event a prior probability of about 0.95.
This probability, together with assuming $b_\psi=0.1$ implies $a_\psi=16$. 

With respect to the MCMC algorithm, we ran two parallel chains, with distant initial starting points, for 
a total of 20,000 iterations for each chain, discarding the first 10,000 iterations for burn-in.  
Thus, inference was based on the last 10,000 iterations from each chain, for a total of 
20,000 iterations.  Convergence has been verified through the Gelman and 
Rubin convergence diagnostics \citep{gelm:rubi:1992}, implemented in the CODA package of 
the R statistical software. Gelman and Rubin convergence diagnostics 
indicated convergence for each of the hyperparameters and for a sample of the 
latent process parameters.

We have applied the model selection approach developed in Section~\ref{sec:model-selection} 
to compare the five models. 
Here we use the first $t^* = 9$ time points as training sample for the computation
of the conditional Bayes factors.
Table \ref{tab:logpd} presents the conditional Bayes factors taking Model~I as the baseline. 
A first feature that is clearly shown by the data is that,  when compared with their spatially 
independent innovations counterparts, models with spatially dependent innovations (SDI)
are better supported by the data. This supports the claim that the long 
incubation period of the HIV \citep{bacc:moss:1989} together with the annual frequency of the
data would lead to spatially correlated field innovations. 
Thus, the data provide evidence of the usefulness of spatially dependent innovations to
account for the fact that an 
increase in the risk of new cases of AIDS in a given year 
may reflect the convolution of HIV infections in several previous years.

\begin{table}[t]
\caption{AIDS Data - Logarithm of conditional Bayes factor (BF) (Baseline: Model I with spatially
dependent innovations). SDI: spatially dependent innovations. 
\label{tab:logpd}}
\begin{center}
\begin{tabular}{c|ccccc} \hline
SDI & \multicolumn{5}{c}{Model} \\
                 &             I         &           II           &            III          &              IV          &            V             \\ 
\hline
Yes          &          0          &         44.7       &      -1474.7     &            79.3        &     \textbf{84.7}          \\
\hline
No            &   -3932.0     &   -1681.2       &      -2809.0     &            26.1        &          26.0           \\
\hline
\end{tabular}
\end{center}
\end{table}

Thus, we now focus on comparison of models with spatially dependent innovations.
With a conditional Bayes factor equal to 84.7, the best is Model V that includes field 
contamination together with a common gradient for all counties.
In addition, Model IV has a conditional Bayes factor somehow lower equal to 79.3. 
Model IV includes a first-order field level and a common gradient for all counties.
Furthermore, Model III that is a second-order field model with a gradient field, that is a gradient 
for each county, has a very small Bayes factor equal to -1474.7. Thus, there is not enough
information in the data to estimate a distinct gradient for each county. 
Therefore, the data provides evidence of the usefulness of the inclusion of both a 
contamination evolution for the log-risk level and a common gradient for all counties.

\begin{table}[t!]
\caption{Posterior summaries for the parameters of Model V that includes spatially dependent 
innovations with contamination field and a common gradient for all counties.
\label{tab:posterior_summaries}}
\centering
\begin{tabular}{c|cc} \hline
Parameter & Mean & 95\% Credible Interval \\ \hline
$\kappa$  &   0.001 &    (0.0005, 0.0014) \\
$\tau$        &   7.16   &    (3.67, 11.92) \\
$\phi$        &   0.50   &    (0.15, 1.14) \\
$\psi$        &   136.7 &    (79.2, 211.2) \\
\hline
\end{tabular}
\end{table}

Table \ref{tab:posterior_summaries} provides posterior means and 95\% credible intervals
for the hyperparameters of the best model, Model V.
A first feature that emerges from the table is that the contamination coefficient $\kappa$ 
is most probably small with 
a posterior mean equal to 0.001 and a 95\% credible interval given by (0.0005, 0.0014).
Even though $\kappa$ seems to be small, the logarithm of the conditional Bayes factor of Model V
against the corresponding model without contamination, that is Model IV, is 
equal to 84.7-79.3 = 5.4 and gives evidence in favor of including the contamination
component in the model. Another feature that emerges is that $\phi$, the parameter
that controls the degree of spatial dependence of the innovations, is also small
with a posterior mean equal to 0.50 and a 95\% credible interval given by (0.15, 1.14).
This small value for $\phi$ has to be put in perspective by computing the conditional
Bayes factor of Model V with spatially dependent against spatially independent 
innovations, that is equal to 84.7-26.0 = 58.7. Hence, the data supports the hypothesis
of spatially dependent innovations. Finally, the precision parameter $\psi$ for the evolution of the
common gradient has a posterior mean equal to 136.7 and a 95\% credible interval given by 
(79.2, 211.2), implying that the common gradient evolves through time with small changes
between subsequent time points. 

\begin{figure}[t!] \centering
\begin{center}
\includegraphics[width=3.0in,height=2.0in]{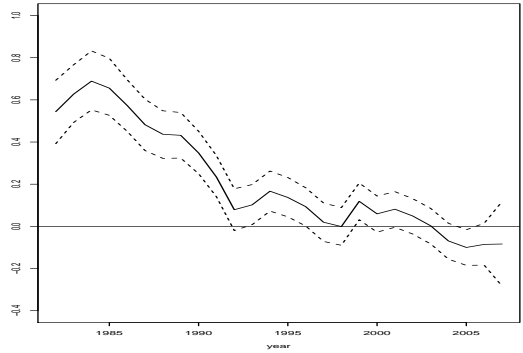}
\end{center}
\caption{AIDS data -  Common gradient for all counties for Model V. Posterior mean (solid line) and
95\% credible interval (dashed lines). } \label{fig:common_gradient}
\end{figure}

Figure~\ref{fig:common_gradient} presents the plot of common gradient of the log-risk
level for all counties through
time for Model V. Specifically, the figure presents the posterior mean (solid line) and 
95\% credible interval (dashed lines). This figure is quite informative, and clearly 
shows that there was a rapid expansion of the epidemic during the 1980's. Even though
this rapid expansion was followed by a sharp reduction in the gradient in the beginning of the 1990's, 
the gradient continued to be small but mostly positive until about 2001. From 2003 to 2007 the
gradient has been mostly negative, and fortunately the trend at the most recently considered year
of 2007 was the reduction in the rate of new cases of AIDS.

\begin{figure} [th!]
\begin{center}
\begin{tabular}{ccc}
Year &  Observed & Fitted   \vspace{-1.0in}\\
\raisebox{21.0ex}{1982}
&
\includegraphics*[width=2.2in,height=3.2in]{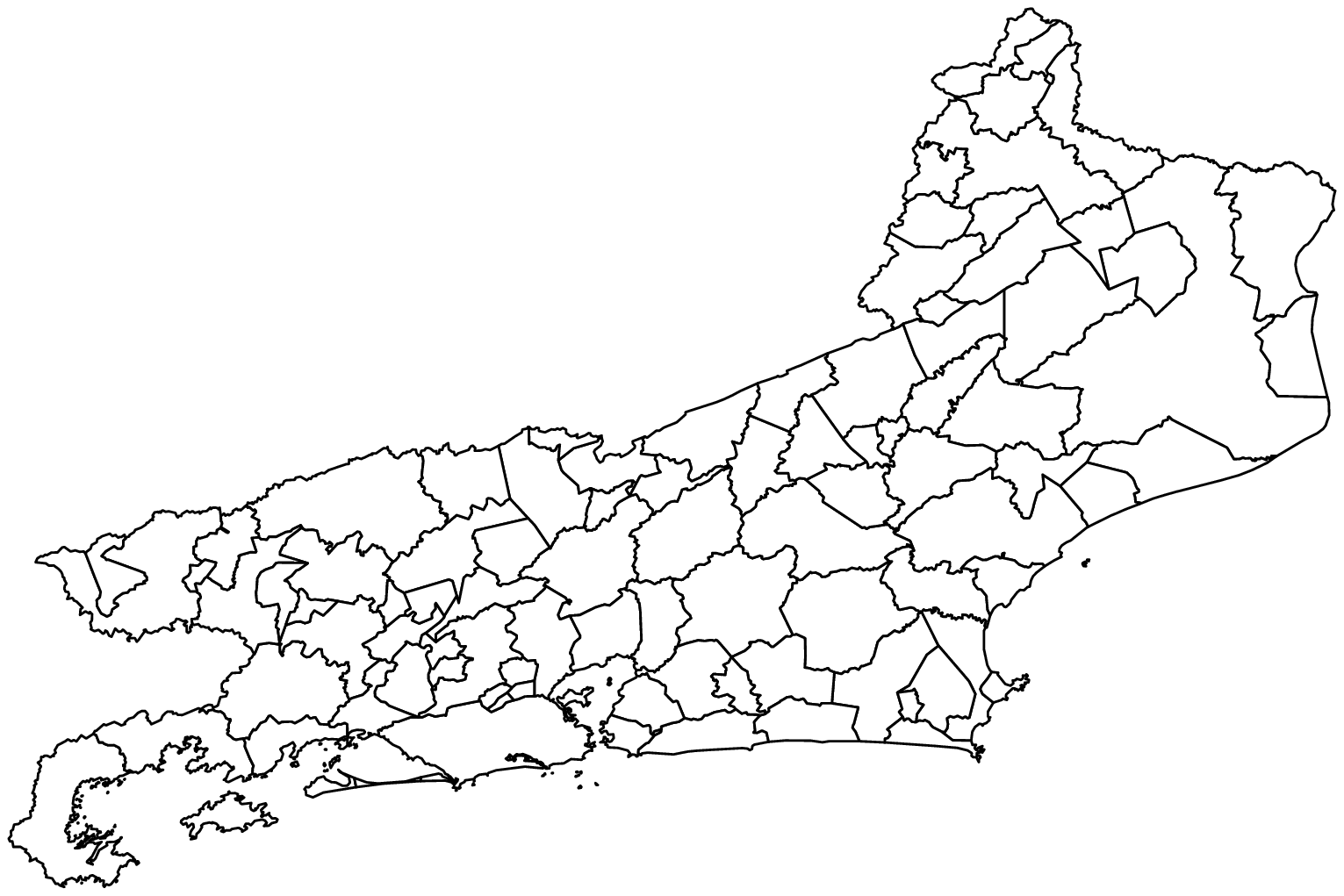}
&
\includegraphics*[width=2.2in,height=3.2in]{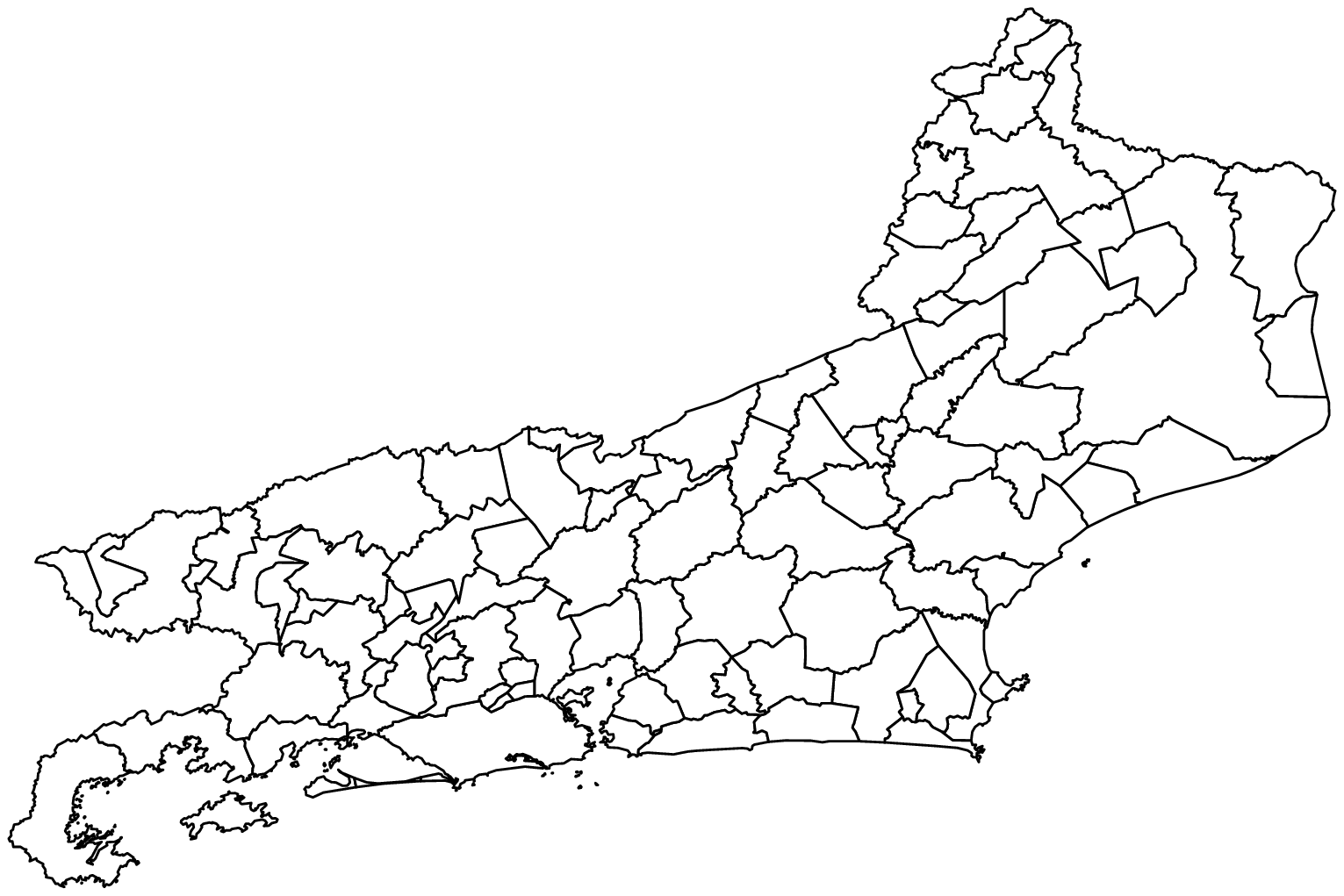}
 \vspace{-2.0in} \\
\raisebox{21.0ex}{1988}
&
\includegraphics*[width=2.2in,height=3.2in]{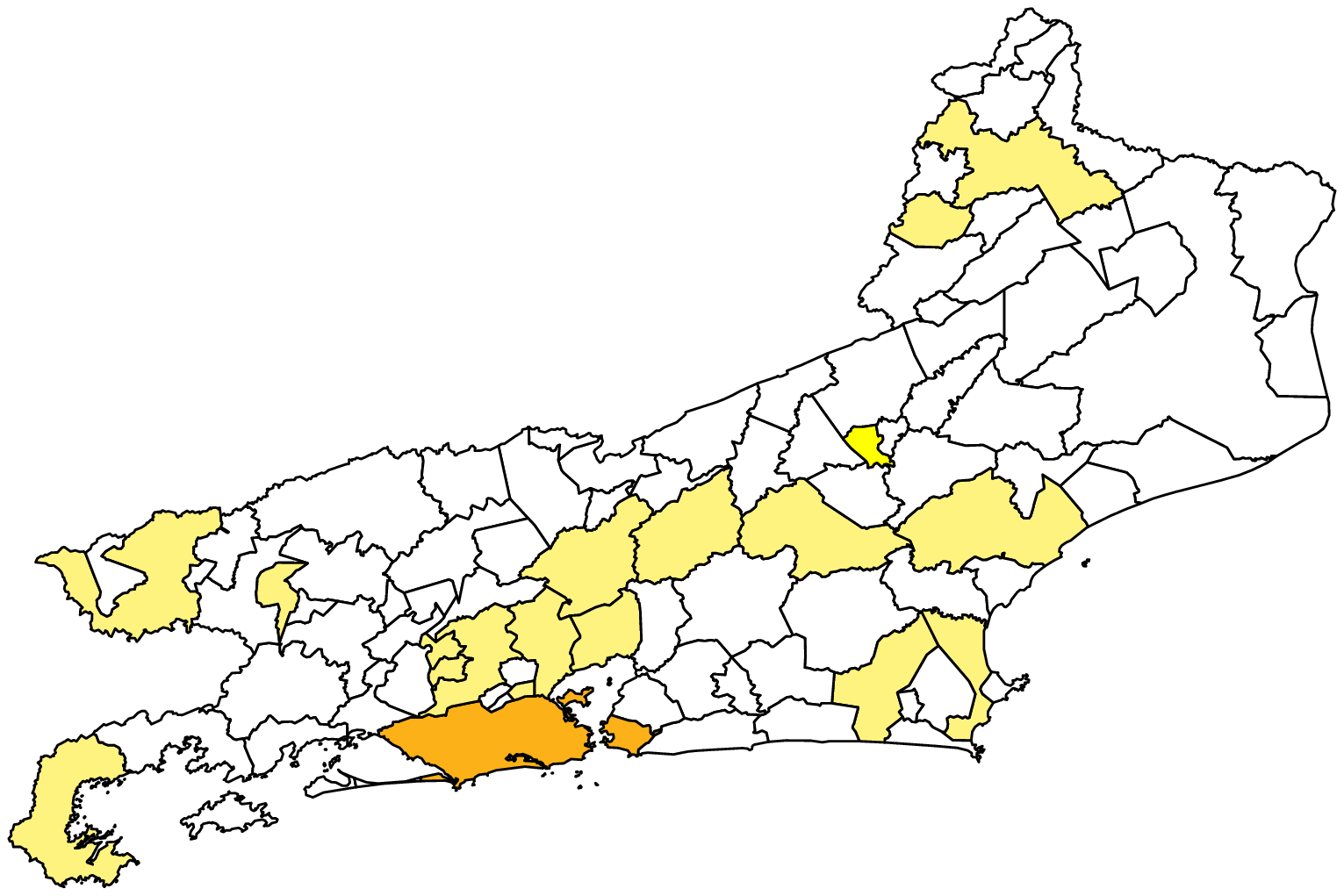}
&
\includegraphics*[width=2.2in,height=3.2in]{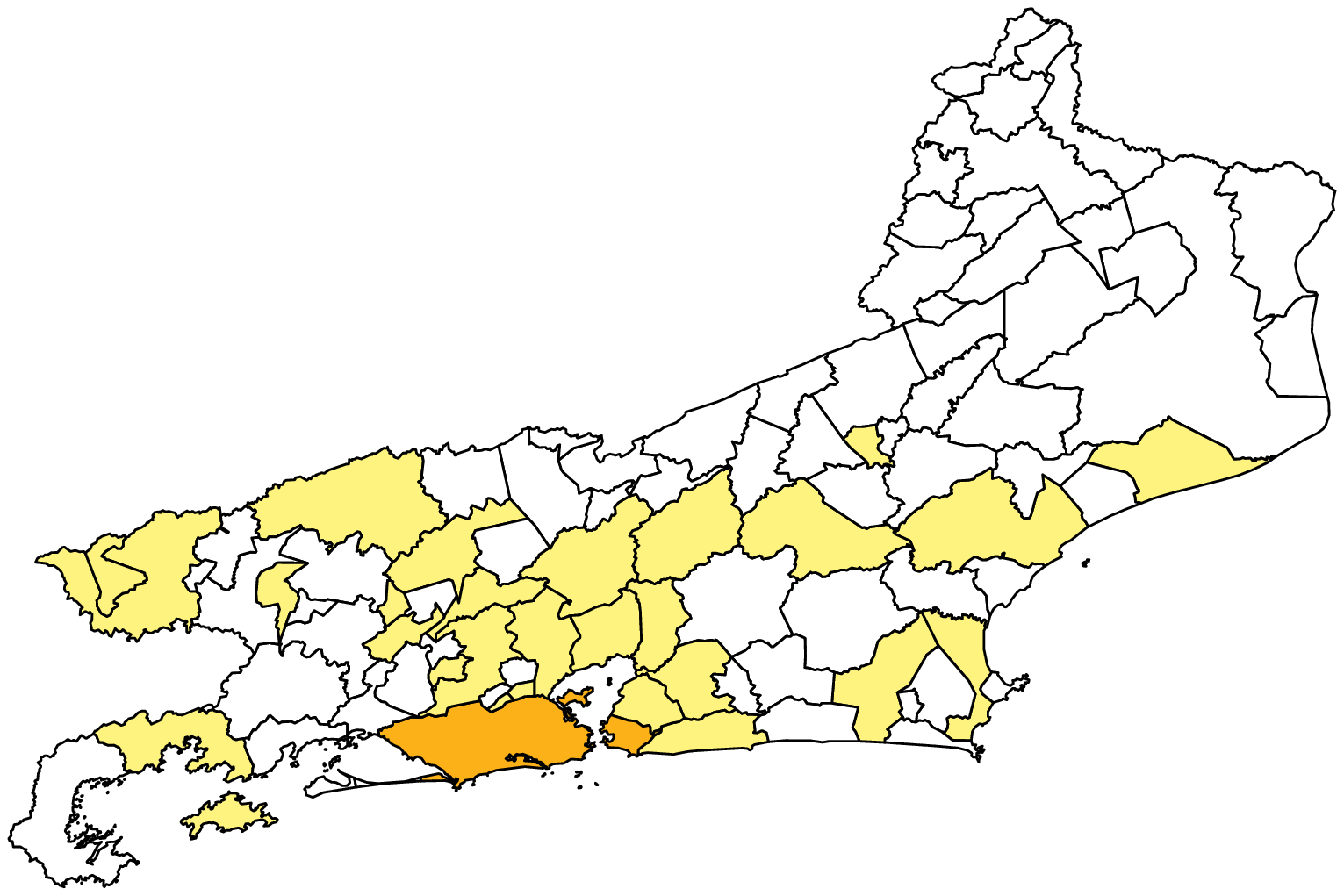}
\vspace{-2.0in} \\
\raisebox{21.0ex}{1994}
&
\includegraphics*[width=2.2in,height=3.2in]{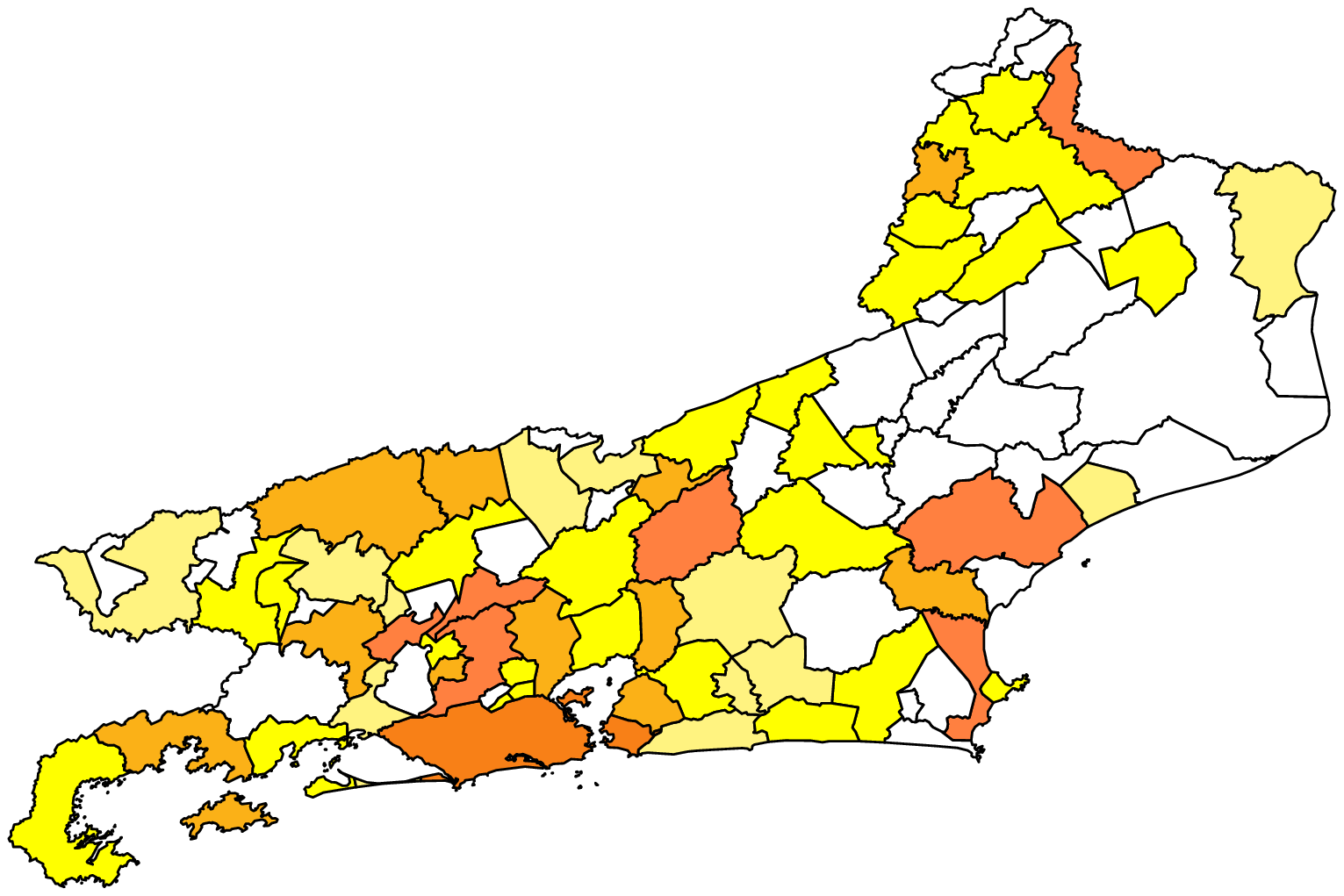}
&
\includegraphics*[width=2.2in,height=3.2in]{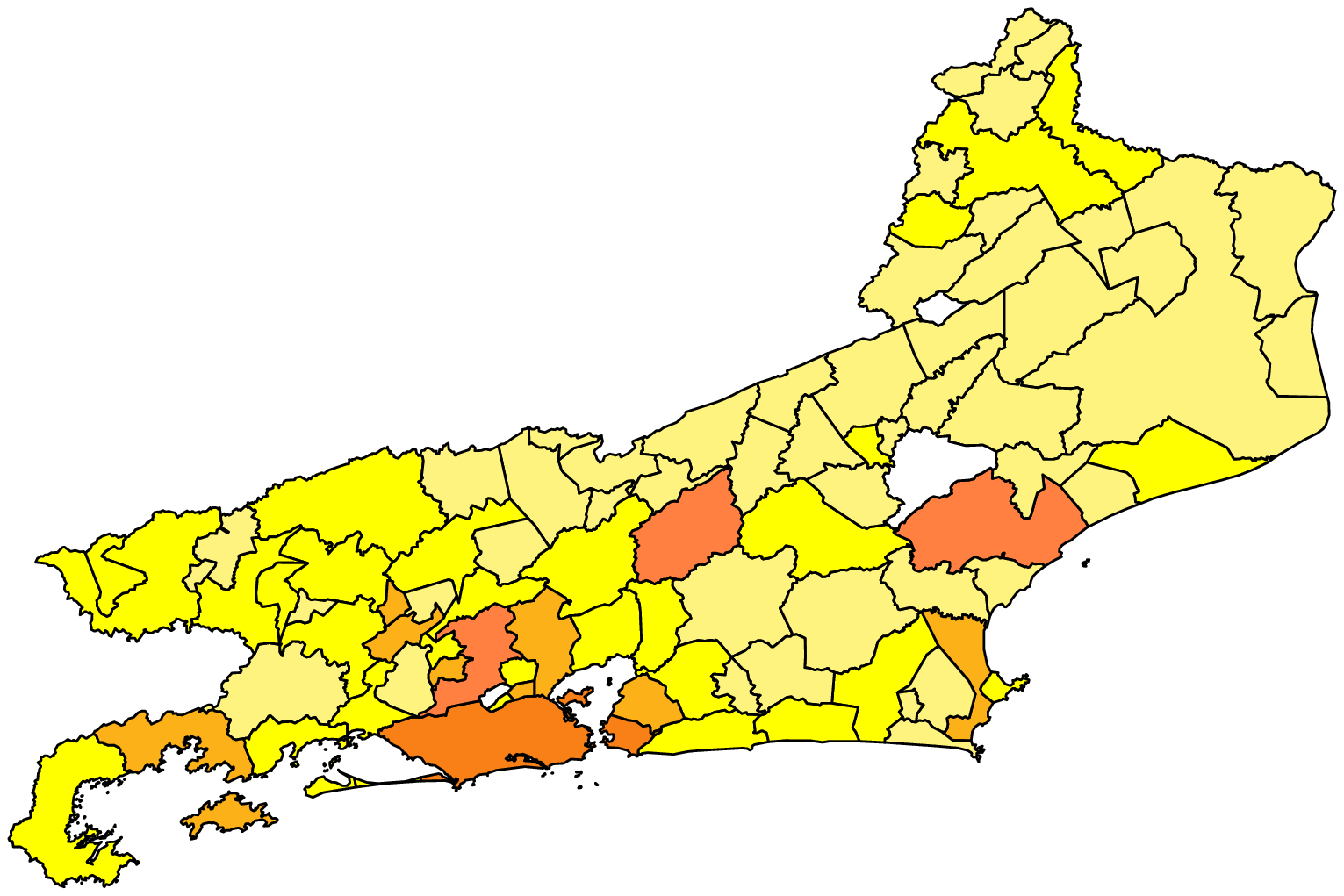}
\vspace{-2.0in} \\
\raisebox{21.0ex}{2000}
&
\includegraphics*[width=2.2in,height=3.2in]{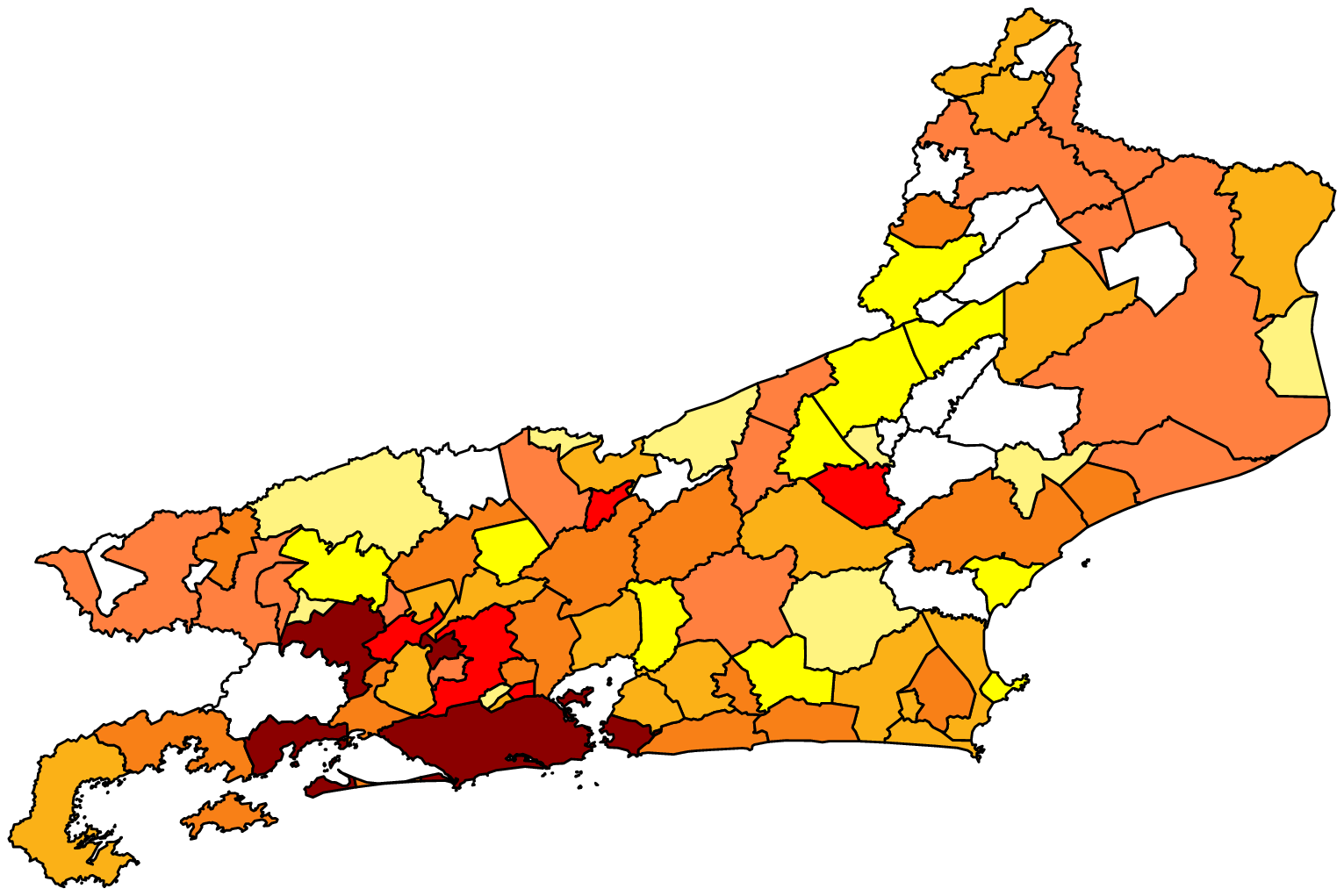}
&
\includegraphics*[width=2.2in,height=3.2in]{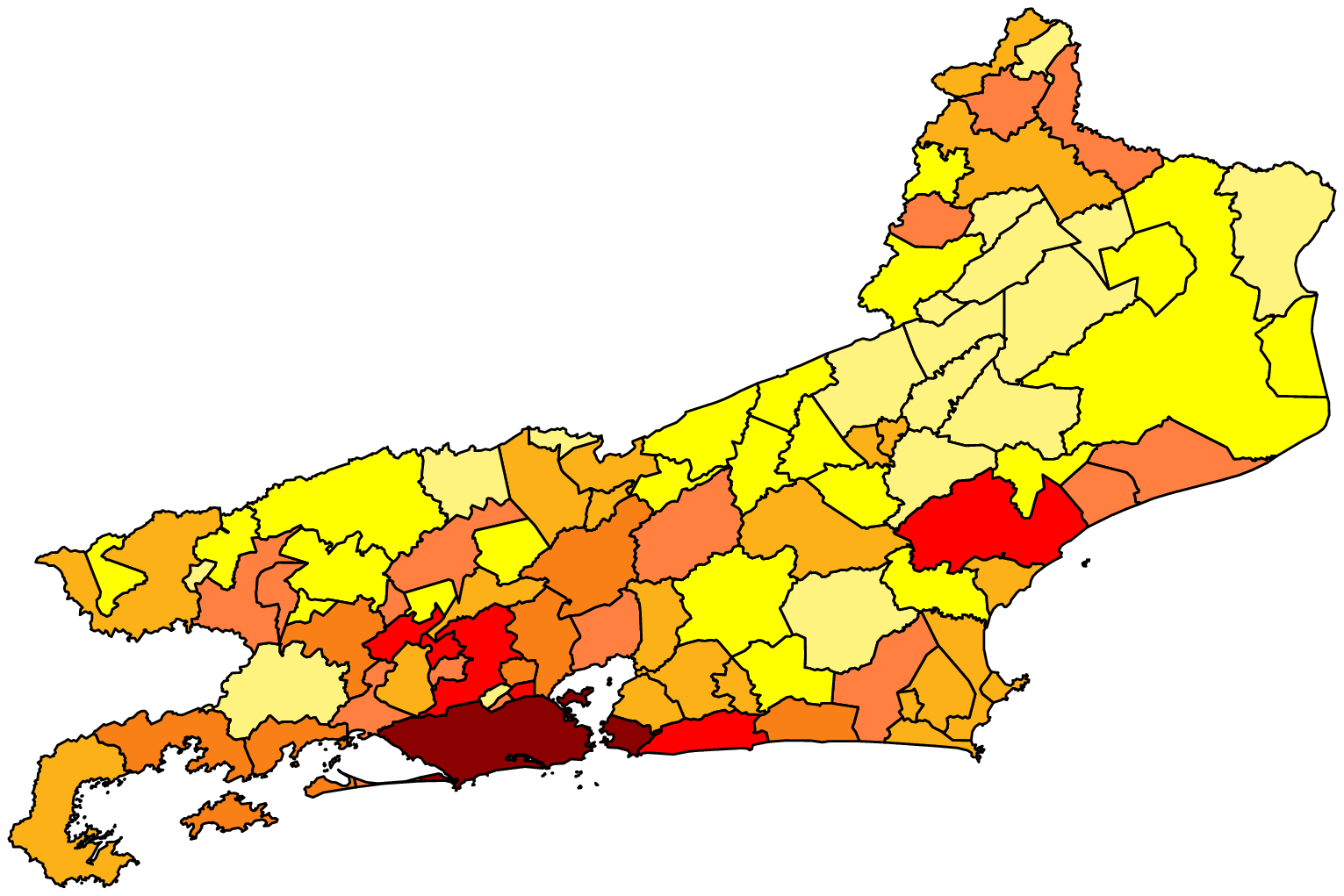}
 \vspace{-2.0in} \\
\raisebox{21.0ex}{2006}
&
\includegraphics*[width=2.2in,height=3.2in]{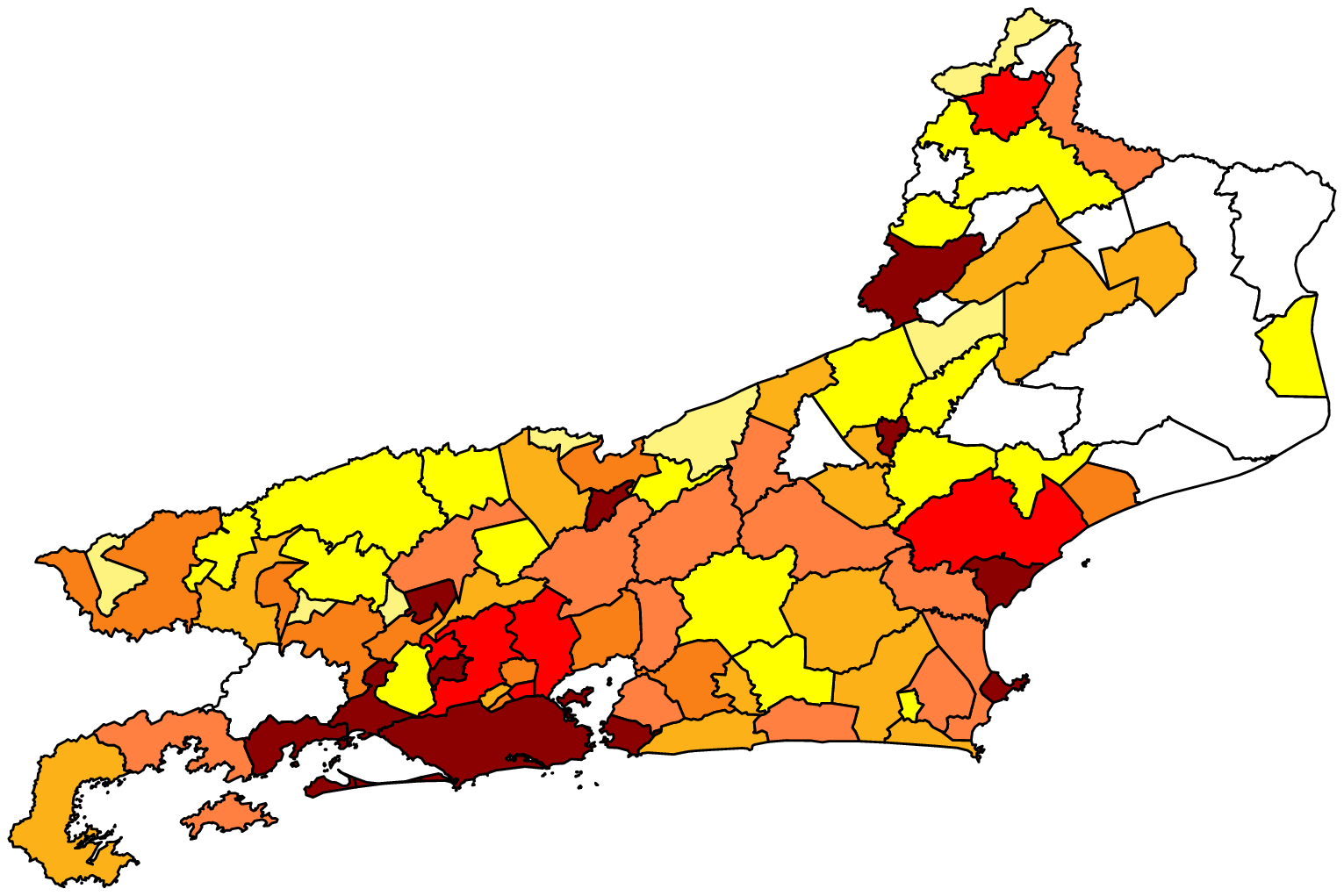}
&
\includegraphics*[width=2.2in,height=3.2in]{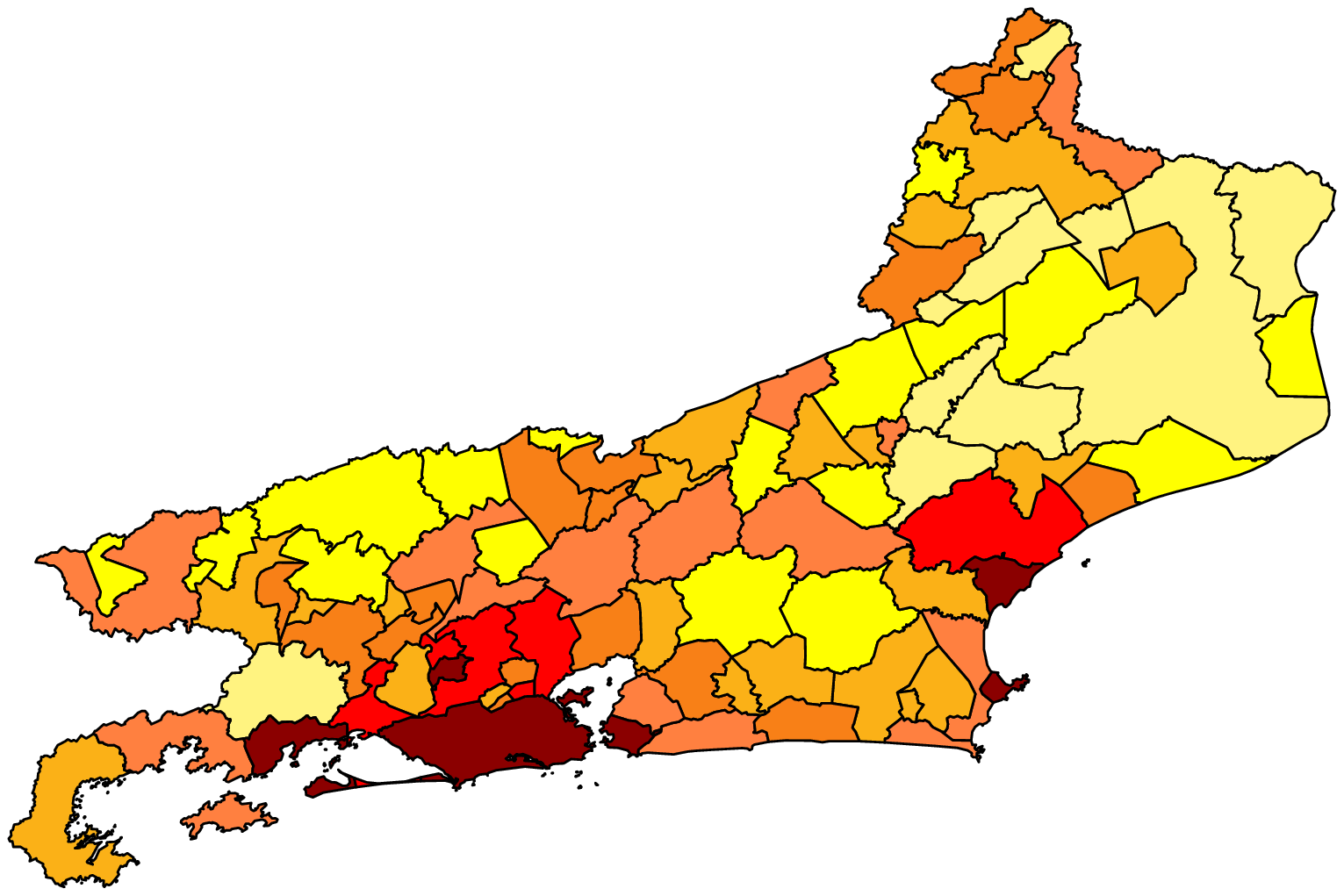}
\vspace{-1.8in}  \\
&
\multicolumn{2}{c}{\includegraphics*[width=3.0in,height=3.0in]{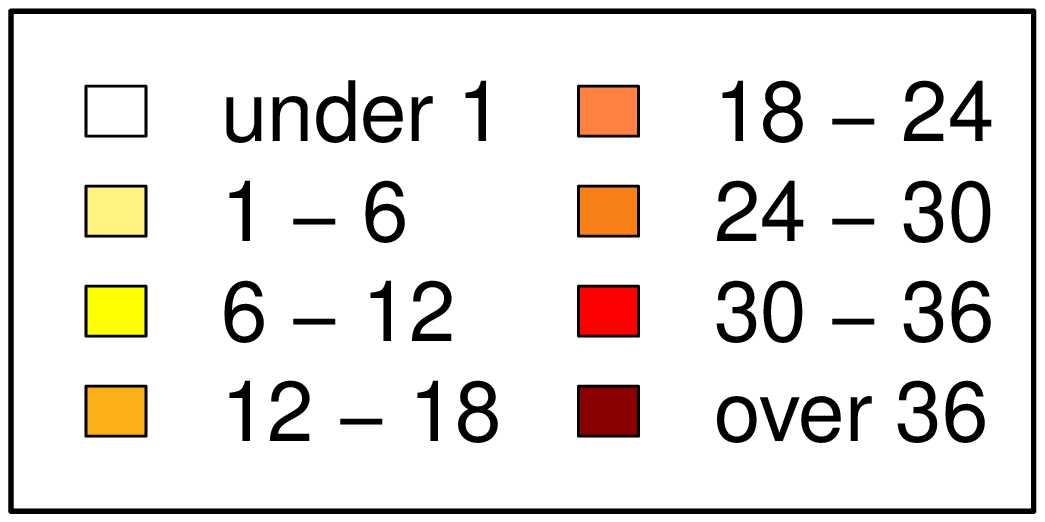}}
\vspace{-1.2in}
\end{tabular}
\end{center}
\caption{AIDS data - Model V. Standardized ratio of new cases per 
100,000 inhabitants: observed (left panels), and fitted (right panels).}
\label{fig:AIDS_obs_and_fitted}
\end{figure}

Figure \ref{fig:AIDS_obs_and_fitted} shows for Model V the observed standardized 
ratio of new cases (left) and the posterior mean of the risk (right) per 100,000 inhabitants 
for the years 1982, 1988, 1994,  2000 and 2006. The maps show the level of risk 
in a scale from white (low risk) to dark red (high risk). 
From a methodological perspective, our procedure accomplishes a nice balance
between spatial and temporal smoothing resulting from the fact that Model V
takes into account the information from spatial neighbors and from across-time
neighbors. 
From an epidemiological perspective, two relevant features emerge from 
Figure~\ref{fig:AIDS_obs_and_fitted}. First, corroborating the findings from 
Figure~\ref{fig:common_gradient}, the rate of new cases increased substantially
from the beginning of the 1980's to the end of the 1990's, and has somehow stabilized
since 2000. Second, the rate of new cases is substantially higher in the 
metropolitan region of the City of Rio de Janeiro, located in the center-south of 
Rio de Janeiro State. These findings may be used by the Department of Health 
for allocating resources both for treatment and prevention of AIDS.

\section{Conclusions}\label{sec:conclusions}

We have presented a novel class of spatiotemporal models for Poisson areal 
data that are useful for the statistical modeling of epidemics. Further, we have
developed several specific models for count data
with several distinct specifications of the evolution system equation to allow different
spatiotemporal process behaviors with time-specific spatially correlated
innovations. In addition, we have developed simulation-based full Bayesian analysis
for both parameter estimation and model selection.

We have applied our spatiotemporal framework to analyze the number of new cases
of AIDS per county in the State of Rio de Janeiro. 
Our spatiotemporal modeling framework has achieved a nice balance
between spatial and temporal smoothing. In particular, it has become clear from
our analysis that in the period of study the rate of new cases of AIDS was substantially higher in the 
metropolitan region of the City of Rio de Janeiro, and that in 2006 the rate of new
cases seemed to be stabilizing. However, it is important to point out
that these were not good news, because the stabilization of the rate of new cases 
meant that the accumulated number of cases continued to increase over time.


From a modeling point of view, our general class of models may be tailored to specific applications.
For example, when monitoring a particular disease, information on the natural history of the disease 
such as incubation time and rate of infection could potentially be incorporated in the spatiotemporal 
model. In one possible specification, parameters of the spatiotemporal model could be assumed to be
functions of parameters of the natural history. In an alternative specification, the natural history 
parameters could be incorporated in the priors for the spatiotemporal model parameters.
Finally, even though we have focused here on the analysis of Poisson data, 
our  methodology is general enough to be applied to observations
with any distribution in the regular exponential family. 
This may be useful for the analysis of spatiotemporal gamma-distributed survival data,
and binomial-distributed survey data.

{\bf DISCLAIMER:} 
Although Juan C. Vivar is an FDA/CTP employee, this work was not done as part of his official duties. This publication reflects the views of the authors and should not be construed to reflect the FDA/CTP's views or policies.

\bibliographystyle{chicago}
\bibliography{biblio}

\end{document}